\documentclass[sigconf]{acmart}

\usepackage{enumitem}
\usepackage{multirow}
\usepackage{colortbl}
\usepackage{diagbox}
\usepackage{color}
\usepackage{rotating}
\usepackage{hyperref}
\usepackage{balance}
\usepackage{booktabs}
\usepackage{algorithm}
\usepackage{subcaption}
\usepackage{algpseudocode}
\usepackage[normalem]{ulem}
\useunder{\uline}{\ul}{}

\AtBeginDocument{%
  }

\setcopyright{acmlicensed}
\copyrightyear{2018}
\acmYear{2018}
\acmDOI{XXXXXXX.XXXXXXX}
\acmConference[Conference acronym 'XX]{Make sure to enter the correct
  conference title from your rights confirmation email}{June 03--05,
  2018}{Woodstock, NY}
\acmISBN{978-1-4503-XXXX-X/2018/06}

\begin{document}

\title{TimeRoute: Time-Aware Modality Routing and Diffusion for Multi-Modal Recommendation}

\author{Pengyu Zhang}
\email{p.zhang@uva.nl}
\orcid{0000-0001-5111-4487}
\affiliation{
  \institution{University of Amsterdam}
  \city{Amsterdam}
  \country{The Netherlands}
}

\author{Yangqin Jiang}
\orcid{0009-0002-3936-1703}
\affiliation{
  \institution{University of Hong Kong}
  \city{Hong Kong}
  \country{China}
}

\author{Klim Zaporojets}
\orcid{0000-0003-4988-978X}
\affiliation{
  \institution{Aarhus University}
  \city{Aarhus}
  \country{Denmark}
}

\author{Congfeng Cao}
\orcid{0000-0001-9011-3807}
\affiliation{
  \institution{University of Amsterdam}
  \city{Amsterdam}
  \country{The Netherlands}
}

\author{Paul Groth}
\orcid{0000-0003-0183-6910}
\affiliation{
  \institution{University of Amsterdam}
  \city{Amsterdam}
  \country{The Netherlands}
}

\renewcommand{\shortauthors}{Anonymous Author(s)}

\begin{abstract}
Multi-modal recommenders fuse user-item interaction signals with item modalities such as text, images, and audio, but the usefulness of each drifts over time and at different rates. For example, around Valentine's Day, chocolate purchases become less driven by textual ingredient cues and more by visual packaging and ambient audio. This \emph{modality time-scale mismatch} gives rise to two coupled challenges: (1) users with different temporal behavior profiles require different modality proportions, and (2) less relevant modalities are more likely to introduce outdated or misleading signals into the recommender. We address both challenges within a unified diffusion-based recommender, \textbf{TimeRoute}. A temporal-aware modal router maps each user's aggregated temporal profile to a personalized modality distribution, replacing the globally shared fusion weights used in prior work. The diffusion-based graph reconstructor is conditioned on the same profile through Feature-wise Linear Modulation (FiLM) with dual-stream long- and short-term denoising heads. This design captures both slowly and rapidly evolving temporal dynamics to suppress outdated modality edges before they enter the propagation graph. Experiments on TikTok, Amazon-Baby, and Amazon-Sports, averaged over 10 seeds, demonstrate consistent improvements over strong baselines across Recall@K, Precision@K, and NDCG@K, reaching up to 9.8\% (P@20 on Amazon-Baby). Controlled attribution studies further show that these gains require both the proposed mechanisms and temporal input: naively granting the backbone the same temporal profile yields no benefit, and feeding the router random noise performs no better than removing the router entirely. Code is available at \url{https://anonymous.4open.science/r/TimeRoute}.
\end{abstract}

\ccsdesc[500]{Information systems~Recommender systems}
\ccsdesc[300]{Computing methodologies~Neural networks}
\ccsdesc[100]{Computing methodologies~Probabilistic reasoning}

\keywords{Multi-modal Recommendation, Time-aware Recommendation, Diffusion Models, Modality Fusion}


\maketitle

\section{Introduction}
\label{sec:intro}

Multi-modal recommender systems power large-scale platforms such as e-commerce marketplaces and short-video applications, where rich item content from images, text, and audio helps overcome sparse user-item feedback~\cite{10.1145/3696410.3714676,10.1145/3664647.3681498}. Recent graph-based approaches that fuse modality-specific signals with collaborative filtering through graph neural networks achieve strong performance across benchmarks~\cite{10.1145/3394171.3413556}.

\begin{figure}[t]
\centering
\includegraphics[width=\linewidth]{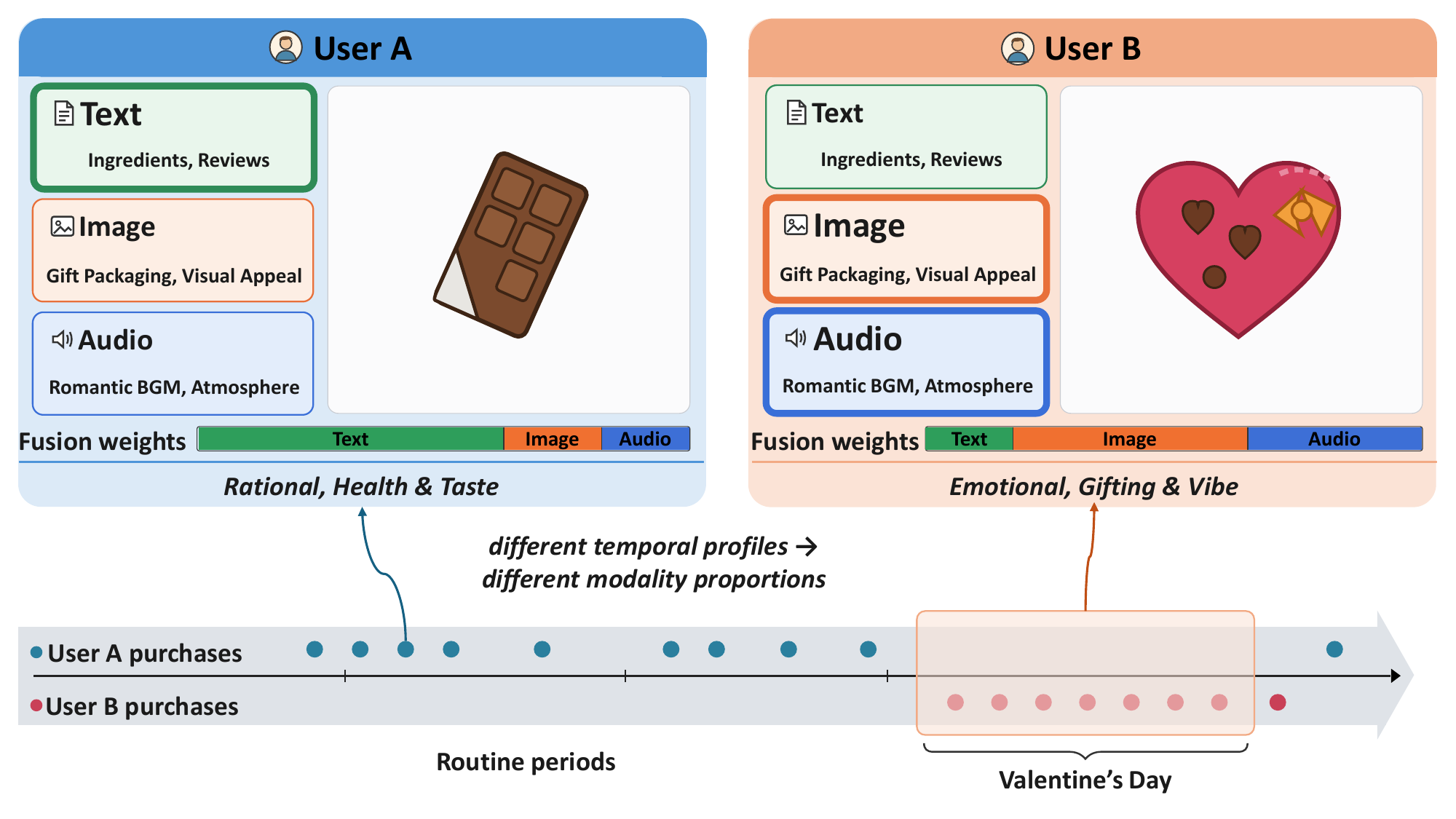}
\caption[Modality time-scale mismatch across user temporal profiles]{An example of modality time-scale mismatch. Around Valentine's Day, the relevance of chocolate purchase modalities shifts from textual cues toward visual and audio signals. User~A's purchases are spread across routine periods, while User~B's are concentrated in the holiday window; the two users therefore require different modality proportions, which a globally shared fusion weight cannot capture.}
\label{fig:modality_timescale_mismatch}
\Description{Two side-by-side panels compare User A and User B. In the User A panel, the Text modality block (ingredients, reviews) is highlighted while the Image and Audio blocks are grayed out, and a fusion-weight bar is dominated by Text. In the User B panel, the Image (gift packaging) and Audio (romantic background music) blocks are highlighted while Text is grayed out, and the fusion-weight bar is dominated by Image and Audio. A timeline below both panels marks a highlighted Valentine's Day window: User A's purchase dots are spread across routine periods, while User B's purchase dots are clustered inside the holiday window. An annotation reads that different temporal profiles lead to different modality proportions.}
\end{figure}

However, the usefulness of each modality is not static. User interests and item popularity drift over time~\cite{10.5555/3600270.3600710}, and the drift speed varies across modalities~\cite{zhu2025probabilisticframeworktemporaldistribution,zheng2025retrievaldistilltemporaldata}. We refer to this phenomenon as \emph{modality time-scale mismatch}: the relevance of each modality shifts over time, but not uniformly across modalities. Figure~\ref{fig:modality_timescale_mismatch} illustrates this point. Chocolate is purchased year-round, but around Valentine's Day purchase decisions shift from textual cues such as ingredients toward visual packaging and atmospheric audio, then revert after the holiday. The key observation is that different modalities become more or less relevant at different moments and evolve at different rates~\cite{10.1145/3746252.3761390}. Users whose interaction histories are situated differently relative to these shifts, in the holiday window for one user and in routine periods for another, therefore need different modality proportions.

This mismatch gives rise to two interrelated problems. The first problem (\textbf{Problem~1}) is that users in different temporal contexts require different modality fusion proportions. Since modality importance shifts at different rates, the appropriate fusion depends on the temporal context reflected in a user's behavior: interaction frequency, recency, and distribution encode a user's position relative to these shifts~\cite{siraudin2025cometh}. However, existing methods, including DiffMM~\cite{10.1145/3664647.3681498}, rely on globally shared fusion weights, assigning the same modality mixture to a user whose recent history is dominated by a holiday shopping spree and to one whose history reflects routine purchases spread over months.

To address this, we introduce a \emph{temporal-aware modal router} that infers each user's temporal context from aggregated behavioral features and produces a per-user softmax distribution over modalities. The router output determines how modality-specific embeddings are fused before GCN propagation, enabling the model to emphasize the modalities most relevant to each user's temporal behavior profile, rather than applying one globally shared mixture to all users. The router personalizes fusion \emph{across users}, assigning one fixed distribution to each user based on their aggregated temporal profile rather than re-routing at each prediction moment (see Limitations).

The second problem (\textbf{Problem~2}) is the flip side of the first: \emph{a modality that is highly relevant in one temporal context can become a source of noise in another}. Signals that were once informative may persist long after they have lost their relevance~\cite{10.1145/3539618.3591663,liu2025preference}; for example, Valentine-themed visual cues may remain in the graph shortly after Valentine's Day, even when they are no longer useful for routine purchases. Modalities that are less relevant in the current context are more likely to contribute such outdated signals. While the modal router partially mitigates this by down-weighting less-relevant modalities, the second problem also manifests at the graph level, where modality-induced edges derived from outdated signals propagate noise through message passing.

To address this structural aspect, we condition the diffusion-based graph reconstruction~\cite{10.1145/3664647.3681498} on temporal signals through Feature-wise Linear Modulation (FiLM)~\cite{10.5555/3504035.3504518}, which applies learned feature-wise scale and shift parameters to modulate the user context. We combine this conditioning with dual-stream long- and short-term denoising heads~\cite{zhang2025timeseriesanalysisfrequency,10.1145/3711896.3736571}. The reconstructor thus accounts for temporal context when rebuilding modality-specific graphs, with the aim of suppressing outdated edges before they propagate noisy signals downstream. Together with the modal router, the framework provides temporal awareness at two complementary levels: the reconstruction of modality-specific graphs (Problem~2) and the fusion of their resulting embeddings (Problem~1).

We evaluate TimeRoute on TikTok, Amazon-Baby, and Amazon-Sports under a 10-seed protocol, and summarize our contributions as follows:
\begin{itemize}[leftmargin=*]
    \item We identify \emph{modality time-scale mismatch} as the root cause of two interrelated problems: (1)~users with different temporal behavior profiles need different modality fusion proportions, and (2)~modalities that are less relevant in the current context are more likely to carry outdated, misleading signals.
    \item We propose TimeRoute, in which a temporal-aware modal router personalizes modality fusion per user (Problem~1) and a time-conditioned diffusion reconstructor uses FiLM modulation with dual-stream long- and short-term heads to suppress outdated modality edges (Problem~2).
    \item Beyond consistent improvements over strong baselines on three benchmarks, controlled attribution studies test the source of these gains: injecting the same temporal profile without our mechanisms yields no benefit, and feeding the router random noise performs no better than removing the router entirely; quartile analysis further shows the learned routing is systematically user-specific.
\end{itemize}

\section{Related Work}
\label{sec:related_work}

\paragraph{Graph-based Multi-Modal Recommendation}
Multi-modal recommendation leverages item content such as images, text, and audio to alleviate interaction sparsity. VBPR~\cite{10.5555/3015812.3015834} pioneered the use of visual features in ranking models, and ACF~\cite{10.1145/3077136.3080797} introduced attention-based reweighting across modalities. Recent work adopts graph neural networks as the backbone: MMGCN~\cite{10.1145/3343031.3351034} and GRCN~\cite{10.1145/3394171.3413556} propagate modality signals over user-item graphs, LATTICE~\cite{10.1145/3474085.3475259} mines latent content-induced structures, and BM3~\cite{10.1145/3543507.3583251} adds bootstrap-style self-supervision. Across this line, however, modality fusion is treated as time-invariant: a single set of fusion weights is learned and applied uniformly across users. This is at odds with the modality time-scale mismatch identified in Section~\ref{sec:intro}, where users with different temporal behavior profiles require different fusion weights. Our temporal-aware modal router instead derives user-adaptive fusion weights from each user's temporal behavior profile, replacing the globally shared mixture.

\paragraph{Temporal Modeling for Recommendation}
Koren's seminal work~\cite{10.1145/1721654.1721677} modeled temporal drift in collaborative filtering, and sequential recommenders such as TiSASRec~\cite{10.1145/3336191.3371786} inject time-interval information into self-attention; we compare against TiSASRec empirically under the chronological split (Appendix). More recent work addresses temporal distribution shift in deployed systems~\cite{zhu2025probabilisticframeworktemporaldistribution,zheng2025retrievaldistilltemporaldata}. In multi-modal settings, XSMoE~\cite{10.1145/3746252.3761390} handles streaming distribution shift with expandable side experts that absorb newly arriving modalities, and HM4SR~\cite{10.1145/3696410.3714676} is most closely related to our motivation: it exploits explicit timestamp signals through a hierarchical time-aware mixture of experts for multi-modal \emph{sequential} recommendation. TimeRoute differs from HM4SR in task setting and mechanism. HM4SR models chronologically ordered interaction sequences and predicts the next item, with experts operating over per-interaction temporal embeddings; TimeRoute addresses non-sequential top-$K$ recommendation, where timestamps enter only through aggregated per-user profiles (Section~\ref{sec:prelim}) that drive fusion-level routing and graph reconstruction. The two are therefore complementary rather than competing designs: one adapts modality use along a sequence, the other personalizes a fixed per-user fusion. More broadly, existing temporal recommenders model user and item dynamics on unimodal sequences or treat prediction as a single evolving signal; TimeRoute instead couples temporal signals with multi-modal fusion itself.

\paragraph{Diffusion Models for Recommendation}
Early generative recommenders used GANs and VAEs~\cite{10.1145/3077136.3080786,10.1145/3178876.3186150}; diffusion models have since been adopted for their training stability, with DiffRec~\cite{10.1145/3539618.3591663} generating interaction vectors through denoising diffusion and subsequent work extending diffusion to sequential recommendation and conditional denoising under contextual signals~\cite{10.1007/978-981-97-2262-4_13,10.1145/3583780.3615134}. In the multi-modal setting, DiffMM~\cite{10.1145/3664647.3681498} reconstructs modality-aware user-item graphs and aligns them via cross-modal contrastive learning, and KDiffE~\cite{11175523} augments diffusion-based multi-modal recommendation with knowledge-enhanced denoising; both serve as our strongest baselines. Closest to our temporal focus, TDPM~\cite{zhu2026tdpm} introduces time-aware diffusion for \emph{generative} recommendation, disentangling long-span period preferences from event-triggered point preferences over semantic-ID tokens. TimeRoute shares the intuition that diffusion should not treat all temporal contexts uniformly, but differs in where time conditioning acts: TDPM conditions token-level generation of next items in a sequential setting, whereas TimeRoute conditions \emph{graph reconstruction}, using FiLM-modulated dual-stream denoisers to suppress outdated modality-induced edges before they enter GCN propagation. In existing multi-modal diffusion recommenders, including DiffMM and KDiffE, the denoiser is driven by a user context that is invariant to temporal behavior; reconstructed graphs therefore cannot reflect when a modality's relevance has changed. Our time-conditioned reconstructor addresses this gap.

\section{Methodology}
\label{sec:method}

\begin{figure*}[thb]
\centering
\includegraphics[width=0.8\linewidth]{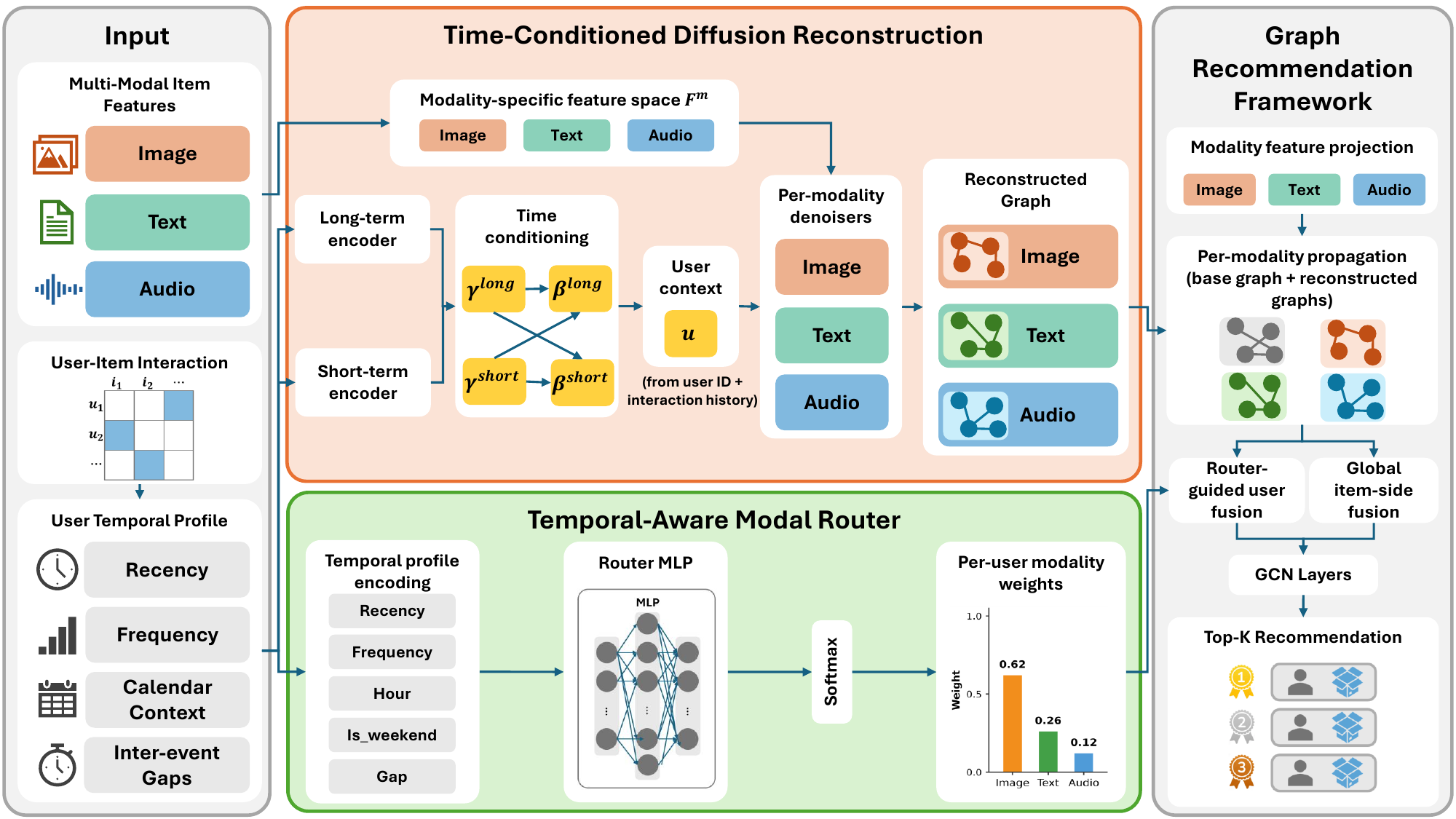}
\caption[Overview of TimeRoute]{Overview of TimeRoute. A per-user temporal profile, built from each user's interaction timestamps, feeds two parallel pathways: the \emph{temporal-aware modal router} (bottom) produces per-user modality fusion weights, and the \emph{time-conditioned diffusion reconstructor} (top) rebuilds modality-specific user-item graphs via dual-stream FiLM modulation of the user context. Both feed the graph recommendation framework (right) for GCN-based top-$K$ recommendation.}
\label{fig:timeroute_overview}
\Description{Architecture diagram of TimeRoute. User interaction timestamps are aggregated into a per-user temporal profile, which feeds two pathways: a temporal-aware modal router that outputs per-user modality fusion weights, and a time-conditioned diffusion reconstructor with dual-stream FiLM heads that rebuilds modality-specific user-item graphs. Both pathways feed a graph recommendation framework that propagates on the base and reconstructed graphs and produces top-K recommendations.}
\end{figure*}

We propose \textbf{TimeRoute} (Figure~\ref{fig:timeroute_overview}), a diffusion-based multi-modal recommender that addresses the two problems identified in Section~\ref{sec:intro}. For \textbf{Problem~1}, we introduce a \emph{temporal-aware modal router}: a per-user adaptive modality weighting mechanism that maps each user's aggregated temporal behavior profile to a personalized distribution over modalities, replacing the globally shared fusion weights used in prior work. The router performs \emph{temporal-profile personalization}: each user receives a fixed fusion distribution derived from their training-set temporal profile. For \textbf{Problem~2}, where modality signals that have lost their relevance persist as outdated edges in the graph, we condition the diffusion-based reconstruction of modality-specific user-item graphs on the same temporal profile, using FiLM modulation~\cite{10.5555/3504035.3504518} with dual-stream long- and short-term heads to suppress outdated modality edges before they enter graph propagation. Together, the two components provide temporal awareness at two complementary levels: the \emph{fusion} of modality-specific embeddings and the \emph{reconstruction} of the graphs that produce them.

\subsection{Preliminaries and Problem Formulation}
\label{sec:prelim}

\noindent\textbf{Task formulation.} We consider implicit-feedback, \emph{non-sequential} top-$K$ recommendation: user preferences are inferred from observed interactions such as clicks, purchases, or views, and the goal is to produce a single ranked top-$K$ list per user. Unlike sequential recommendation, we do not condition predictions on a query timestamp or on the position of the next interaction. Timestamps enter the model only through aggregated per-user temporal profiles (Section~\ref{sec:profile}). Let $\mathcal{U}$ and $\mathcal{V}$ denote the sets of users and items, with interactions encoded by a binary matrix $\mathbf{R} \in \{0,1\}^{|\mathcal{U}|\times|\mathcal{V}|}$, where $R_{u,v}=1$ indicates an observed interaction. We write $\mathbf{r}_u \in \{0,1\}^{|\mathcal{V}|}$ for the $u$-th row of $\mathbf{R}$. We denote the same vector by $\mathbf{x}_0$ in the diffusion formulation in Section~\ref{sec:diffusion}. Following DiffMM~\cite{10.1145/3664647.3681498}, we score user-item pairs by the inner product of their final embeddings, $\hat{y}_{u,v} = \mathbf{h}_u^\top \mathbf{h}_v$, and return the top-$K$ items per user.

\noindent\textbf{Multi-modal inputs.}
Each item $v$ is associated with raw modality features $\hat{\mathbf{f}}^m_v \in \mathbb{R}^{d_m}$ for each modality $m$ in the modality set $\mathcal{M}$. In our experiments, $\mathcal{M}=\{\text{image},\text{text}\}$ on Amazon-Baby and Amazon-Sports, and $\mathcal{M}=\{\text{image},\text{text},\text{audio}\}$ on TikTok.

\noindent\textbf{Temporal inputs.} An observed interaction $(u,v)$ may carry a Unix timestamp $\tau_{u,v}$; coverage varies across datasets (on Amazon-Sports, for example, approximately 63\% of training interactions are timestamped), and profiles are built from the timestamped subset. From user $u$'s timestamped \emph{training-set} interactions, we derive a per-user temporal profile covering a common set of semantic fields (interaction recency and frequency, calendar context, weekday/hour patterns). The two downstream pathways consume this profile in different vectorized forms: the modal router uses a $16$-dimensional form $\tilde{\mathbf{t}}_u \in \mathbb{R}^{16}$ in which cyclic fields are sin-cos encoded (Section~\ref{sec:profile}), and the diffusion conditioner uses a $12$-dimensional raw form $\mathbf{t}_u \in \mathbb{R}^{12}$ processed by per-field encoders (Section~\ref{sec:dualstream}).

\subsection{Graph Recommendation Framework}
\label{sec:pipeline}

We introduce the pipeline before the two main components, so that each component's role in the larger framework is clear.

\noindent\textbf{Modality feature projection and user context.}
Raw modality features are projected into a shared $d$-dimensional space via learnable projectors $\phi_m$ (a linear layer with leaky-ReLU): $\mathbf{f}^m_v = \phi_m(\hat{\mathbf{f}}^m_v) \in \mathbb{R}^{d}$, with the per-item vectors stacked as $\mathbf{F}^m \in \mathbb{R}^{|\mathcal{V}|\times d}$. With learnable ID embedding matrices $\mathbf{E}_U \in \mathbb{R}^{|\mathcal{U}|\times d}$ and $\mathbf{E}_V \in \mathbb{R}^{|\mathcal{V}|\times d}$, we define the \emph{user context embedding}
\begin{equation}
    \mathbf{u} = (\mathbf{E}_U)_u + \mathbf{r}_u\,\mathbf{E}_V \in \mathbb{R}^{d},
    \label{eq:user_u}
\end{equation}
which combines the user's ID embedding with an aggregate of interacted-item embeddings and is the time-modulation target of the FiLM heads (Equation~\eqref{eq:film_apply}).

\noindent\textbf{Modality-specific propagation and fusion.}
The diffusion reconstructor (Section~\ref{sec:diffusion}) produces, for each modality, a reconstructed user-item graph $\widehat{\mathbf{R}}^m$. Following~\cite{10.1145/3664647.3681498}, each modality's embeddings combine two propagation steps: LightGCN-style propagation~\cite{he2020lightgcn} on the base graph $\mathbf{R}$ seeded with the projected modality features yields $\mathbf{e}^{m,\text{base}}_{(\cdot)}$, one-step propagation on $\widehat{\mathbf{R}}^m$ seeded with the ID embeddings yields $\mathbf{e}^{m,\text{rec}}_{(\cdot)}$, and the two are combined as $\mathbf{e}^m_{(\cdot)} = \mathbf{e}^{m,\text{base}}_{(\cdot)} + \lambda_{\mathrm{adj}}\, \mathbf{e}^{m,\text{rec}}_{(\cdot)}$, where $\lambda_{\mathrm{adj}}$ controls the reconstructed graph's contribution. The modality-specific embeddings are then fused on the user side by the temporal-aware modal router (Section~\ref{sec:router}) and on the item side by a globally learned modality weight, yielding $\mathbf{e}^{\mathrm{fuse}}_u$ and $\mathbf{e}^{\mathrm{fuse}}_v$, and a final stack of $L$ GCN layers~\cite{he2020lightgcn} on the base graph, with a normalized residual term controlled by $\lambda_{\mathrm{res}}$, produces the final embeddings $\mathbf{h}_{(\cdot)}$ used for scoring.

\subsection{Temporal-Aware Modal Router}
\label{sec:router}

The modal router is our mechanism for Problem~1: from each user's aggregated temporal profile it produces personalized modality fusion weights, replacing the globally shared mixture. Consistent with Section~\ref{sec:prelim}, the routing is fixed per user and does not re-route at each prediction moment. We describe its five components below; whether this pathway's gains stem from information access or from the mechanism itself is examined in Section~\ref{sec:rq2}.

\subsubsection{User Temporal Profile Construction}
\label{sec:profile}
For each user $u$, we extract $12$ raw temporal fields from the user's timestamped training-set interactions; no validation or test timestamps are used at any stage of profile construction (see the evaluation protocol in Section~\ref{sec:exp_setup}). The fields fall into three semantic groups: \textbf{long-term context} (4 fields, e.g., \textit{global\_days}, \textit{month}), encoding the user's overall temporal position and historical span; \textbf{short-term context} (6 fields, e.g., log-scaled inter-event gap, normalized recency, \textit{weekday}, \textit{hour}), encoding the local temporal structure of recent activity; and \textbf{auxiliary fields} (2 fields), used to disambiguate users with otherwise identical aggregated signatures. Each field is aggregated across the user's interactions using a recency-weighted average, so that more recent interactions contribute more; cyclic fields are aggregated via $\sin$-$\cos$ averaging followed by angular reconstruction, ensuring that, for example, the average of $23{:}00$ and $01{:}00$ is $00{:}00$ rather than $12{:}00$.

\noindent\textbf{Vectorization for the router.} For the router input, the four cyclic fields are converted to $(\sin,\cos)$ pairs to preserve circular proximity, continuous fields are normalized to a comparable range, bounded fields are clipped to $[0,1]$, and the weekend indicator is one-hot encoded, yielding $\tilde{\mathbf{t}}_u \in \mathbb{R}^{16}$. The encoding is lightweight, adding only 610 parameters on Amazon-Baby and Amazon-Sports and 643 on TikTok; the full field list and per-field encoding details are in the Appendix\footnote{\url{https://anonymous.4open.science/r/TimeRoute/TimeRoute_appendix.pdf}} and in our released code.

\subsubsection{Routing Function}
\label{sec:routing_fn}
Given $\tilde{\mathbf{t}}_u$, the router produces a probability distribution over modalities through a two-layer MLP followed by a softmax:
\begin{equation}
    \mathbf{w}_u = \operatorname{Softmax}\!\left(\mathbf{W}_2\, \sigma(\mathbf{W}_1 \tilde{\mathbf{t}}_u + \mathbf{b}_1) + \mathbf{b}_2\right),
    \label{eq:router}
\end{equation}
where $\mathbf{W}_1 \in \mathbb{R}^{32 \times 16}$, $\mathbf{W}_2 \in \mathbb{R}^{|\mathcal{M}| \times 32}$, and $\sigma$ is the ReLU activation. The output $\mathbf{w}_u \in \Delta^{|\mathcal{M}|-1}$ is a per-user modality distribution. Because $\tilde{\mathbf{t}}_u$ aggregates the user's training-set history, $\mathbf{w}_u$ is computed once per user and remains fixed across prediction moments. We initialize $\mathbf{W}_2$ and $\mathbf{b}_2$ to zero so that $\mathbf{w}_u = 1/|\mathcal{M}|$ at initialization: this warm start avoids strongly skewed routing before the modality-specific embeddings have meaningfully differentiated.

\subsubsection{Weight Floor}
\label{sec:floor}
To prevent the router from entirely suppressing a modality, we rescale the softmax output as $w'_{u,m} = \epsilon + (1 - |\mathcal{M}|\epsilon)\, w_{u,m}$ with $\epsilon = 0.05$, which preserves the simplex structure while guaranteeing $w'_{u,m} \geq \epsilon$. This keeps every modality branch contributing a bounded minimum signal and receiving training gradient, reducing the cost of confidently wrong routing decisions.

\subsubsection{Per-User Modality Fusion}
\label{sec:fusion}
Modality fusion is applied \emph{differently} on the user and item sides:
\begin{align}
    \mathbf{e}^{\mathrm{fuse}}_u &= \sum_{m \in \mathcal{M}} w'_{u,m} \cdot \mathbf{e}^m_u, \quad u \in \mathcal{U}, \label{eq:fuse_user} \\
    \mathbf{e}^{\mathrm{fuse}}_v &= \sum_{m \in \mathcal{M}} \beta_m \cdot \mathbf{e}^m_v, \quad v \in \mathcal{V}, \label{eq:fuse_item}
\end{align}
where $\boldsymbol{\beta} = \mathrm{Softmax}(\tilde{\boldsymbol{\beta}})$ is a globally learned modality-weight vector shared across all items. The user side offers a compact temporal profile as router input, whereas an item-side router would multiply routing decisions while resting on much weaker per-item evidence; we therefore retain a global item-side weight $\boldsymbol{\beta}$ for dataset-level modality importance. An item-side router variant degraded performance on all three datasets in screening experiments (see the Appendix).

\subsubsection{Diversity Regularization}
\label{sec:diversity}
The weight floor does not prevent a subtler failure mode: the router can converge to a near-uniform distribution shared across users, recovering DiffMM-like behavior with no per-user adaptation and thus failing to address Problem~1. To explicitly encourage per-user differentiation, we add a diversity regularizer:
\begin{equation}
    \mathcal{L}_{\mathrm{div}} = \underbrace{\frac{1}{|\mathcal{B}|}\sum_{u \in \mathcal{B}} \mathrm{H}(\mathbf{w}_u)}_{\text{per-user entropy}} \;-\; \underbrace{\mathrm{H}\!\left(\frac{1}{|\mathcal{B}|}\sum_{u \in \mathcal{B}} \mathbf{w}_u\right)}_{\text{batch-mean entropy}},
    \label{eq:div_loss}
\end{equation}
where $\mathrm{H}(\mathbf{p}) = -\sum_m p_m \log p_m$ is the Shannon entropy, $\mathcal{B}$ is the current minibatch of users, and $\mathbf{w}_u$ denotes the pre-floor softmax output (Equation~\eqref{eq:router}) to preserve gradient flow through the router during optimization. Minimizing $\mathcal{L}_{\mathrm{div}}$ encourages each user toward a confident routing decision (low per-user entropy) while keeping the batch-averaged distribution broadly balanced across modalities (high batch-mean entropy), so that user-specific distributions differentiate without collapsing to a single globally dominant modality. The regularizer is added to the joint loss with a single hyperparameter $\lambda_{\mathrm{div}}$, tuned per dataset (see our released code at \url{https://anonymous.4open.science/r/TimeRoute} for full values).

\subsection{Time-Conditioned Diffusion Reconstruction}
\label{sec:diffusion}

To address Problem~2, we extend the diffusion-based graph reconstruction pipeline of DiffMM~\cite{10.1145/3664647.3681498} along two axes. \emph{Time conditioning} introduces a FiLM-modulated user context with dual-stream long- and short-term heads (Section~\ref{sec:dualstream}), together with a time-reweighted reconstruction loss (Section~\ref{sec:reweight}). \emph{Modality grounding} adds an alignment regularizer that ties modality-specific reconstructions to the modality feature space (Section~\ref{sec:align}). FiLM conditioning, the dual-stream heads, and the time-reweighted loss modulate the same reconstruction pathway from complementary angles and are designed to act jointly.

We follow the standard DDPM formulation of Ho et al.~\cite{10.5555/3495724.3496298}, as instantiated for recommendation by DiffMM~\cite{10.1145/3664647.3681498}: the forward process gradually corrupts the interaction vector $\mathbf{x}_0 = \mathbf{r}_u$ with Gaussian noise under a fixed variance schedule, and a per-modality MLP denoiser $D_{\theta_m}$ predicts the clean vector $\hat{\mathbf{x}}_0^m = D_{\theta_m}(\mathbf{x}_t, t, \mathbf{u}')$, where $\mathbf{u}'$ is the time-modulated user context, instantiated by the long- and short-term streams of Section~\ref{sec:dualstream}. The reconstructed graphs $\widehat{\mathbf{R}}^m$ assembled from these denoised predictions serve as inputs to the modality-specific propagation of Section~\ref{sec:pipeline}.

\subsubsection{Time Conditioning via FiLM and Dual-Stream Heads}
\label{sec:dualstream}
Each modality $m \in \mathcal{M}$ has its own denoiser $D_{\theta_m}$, with independently learned field-level encoders, fusion MLPs, FiLM heads, and a gating network; we describe the architecture generically. A standard diffusion-based recommender conditions the denoiser on a single user context $\mathbf{u}$ that is identical across temporal contexts; as a result, the reconstruction cannot reflect shifts in the user's recent behavior (Problem~2). We therefore time-modulate $\mathbf{u}$ via FiLM and equip the denoiser with separate long- and short-term streams, designed to capture slowly and rapidly evolving behavioral dynamics, respectively.

\noindent\textbf{Field-level temporal encoding.} Each of the $12$ fields in $\mathbf{t}_u$ is encoded individually (linear layers for continuous fields, $(\sin,\cos)$ pairs for cyclic fields, and an embedding table for a quantile-based inter-event gap bucket), then partitioned by temporal scope: the four slowly evolving fields (overall usage history, seasonal position) feed a long-term stream, and the six rapidly changing fields (inter-event timing, time-of-day and weekday effects, recency) feed a short-term stream. Stream-specific MLPs map the concatenated encodings to hidden states $\mathbf{h}^{\mathrm{long}}_u$ and $\mathbf{h}^{\mathrm{short}}_u$; the exact field-to-stream assignment is listed in the Appendix.

\noindent\textbf{Dual FiLM parameters.} Each stream produces its own scale and shift parameters that modulate the user context $\mathbf{u}$ of Equation~\eqref{eq:user_u}. For stream $s \in \{\mathrm{long}, \mathrm{short}\}$:
\begin{align}
    (\boldsymbol{\gamma}^s_u,\, \boldsymbol{\beta}^s_u) &= \mathcal{F}_s(\mathbf{h}^s_u), \label{eq:film_heads}\\
    \mathbf{u}^s &= \mathbf{u} \odot \boldsymbol{\gamma}^s_u + \boldsymbol{\beta}^s_u, \label{eq:film_apply}
\end{align}
where $\mathcal{F}_{\mathrm{long}}$ and $\mathcal{F}_{\mathrm{short}}$ are linear heads.\footnote{The FiLM shifts $\boldsymbol{\beta}^s_u$ are distinct from the item-side fusion weights $\beta_m$ in Equation~\eqref{eq:fuse_item}; we keep the standard FiLM notation despite the overlap.} Following Perez et al.~\cite{10.5555/3504035.3504518}, we parameterize $\boldsymbol{\gamma} = \mathbf{1} + \mathrm{ELU}(\cdot)$ and $\boldsymbol{\beta} = \tanh(\cdot)$ so that the modulation starts near an identity transformation, which stabilizes early training.

\noindent\textbf{Dual-stream denoising and gated combination.} The denoiser is run twice per training step, once conditioned on $\mathbf{u}^{\mathrm{long}}$ and once on $\mathbf{u}^{\mathrm{short}}$, and the two reconstructions are combined by a user- and step-dependent gate $\alpha_{u,t} = \sigma(\mathrm{MLP}_{\mathrm{gate}}([\,\mathbf{h}^{\mathrm{long}}_u,\, \mathbf{h}^{\mathrm{short}}_u,\, \mathbf{e}_t\,]))$, where $\sigma$ is the sigmoid function and $\mathbf{e}_t$ is the diffusion-step embedding:
\begin{equation}
    \hat{\mathbf{x}}_0^m = \alpha_{u,t} \cdot \hat{\mathbf{x}}_0^{m,\mathrm{long}} + (1 - \alpha_{u,t}) \cdot \hat{\mathbf{x}}_0^{m,\mathrm{short}}. \label{eq:dual_combine}
\end{equation}
The gate interpolates, per user and per diffusion step, between slowly evolving long-term patterns and rapidly changing short-term behavior.

\subsubsection{Modality-aware Alignment Regularization}
\label{sec:align}
The denoiser does not take the modality feature matrix $\mathbf{F}^m$ as input directly. To inject modality semantics, we tie the reconstructed user-item scores to the modality feature space. For user $u$, the denoiser predicts a clean interaction vector $\hat{\mathbf{x}}_{0,u}^m$ over items, where the subscript $0$ denotes the denoised diffusion state rather than a user index:
\begin{equation}
    \mathcal{L}^m_{\mathrm{align}} =
    \mathbb{E}_{u}
    \left[
    \bigl\|
    \hat{\mathbf{x}}_{0,u}^m\,\mathbf{F}^m
    -
    \mathbf{x}_{0,u}\,\mathbf{E}_V
    \bigr\|_2^2
    \right].
    \label{eq:align}
\end{equation}
The term $\hat{\mathbf{x}}_{0,u}^m \mathbf{F}^m$ projects the \emph{predicted} interactions into modality-$m$ feature space, while $\mathbf{x}_{0,u}\mathbf{E}_V$ projects the \emph{observed} interactions into the shared ID embedding space. Minimizing their distance keeps each modality-specific reconstruction faithful to observed behavior while respecting the geometry of modality $m$, enabling modality-differentiated graphs $\widehat{\mathbf{R}}^m$ with a compact denoiser architecture.

\subsubsection{Time-Reweighted Reconstruction Loss and Graph Assembly}
\label{sec:reweight}
A standard diffusion reconstruction loss weights users equally regardless of how recently they were active, even though more recent activity is typically more informative of current preferences. We therefore introduce a time-reweighted variant of the per-modality reconstruction loss:
\begin{equation}
    \mathcal{L}^m_{\mathrm{diff}} = \mathbb{E}_{u,\,t,\,\boldsymbol{\epsilon}}\!\left[\, w_{\mathrm{snr}}(t)\cdot \bigl(1 + \lambda_t \cdot \mathrm{pos\_norm}_u\bigr)\cdot\bigl\|\mathbf{x}_0 - \hat{\mathbf{x}}_0^m\bigr\|_2^2\right],
    \label{eq:time_reweight}
\end{equation}
where $w_{\mathrm{snr}}(t)$ is the standard SNR-derived diffusion-step weight, $\mathrm{pos\_norm}_u \in [0,1]$ is the normalized recency of $u$'s most recent \emph{training} interaction, and $\lambda_t$ is a hyperparameter tuned per dataset. The reweighting prioritizes accurate reconstruction near the temporal frontier, where outdated modality signals are most likely to mislead the recommender (the structural side of Problem~2). Once the per-modality denoisers are trained, we sample $\hat{\mathbf{x}}_0^m$ via reverse diffusion under the FiLM conditioning of Section~\ref{sec:dualstream} and assemble $\widehat{\mathbf{R}}^m$ from the top-$k$ predicted items per user, as illustrated in the rightmost panel of Figure~\ref{fig:timeroute_overview}.

\subsection{Training Objective and Optimization}
\label{sec:training}

The model is optimized with an alternating three-phase schedule under a composite objective:
\begin{equation}
    \mathcal{L} = \mathcal{L}_{\mathrm{BPR}} + \lambda_{\mathrm{cl}}\mathcal{L}_{\mathrm{CL}} + \!\sum_{m \in \mathcal{M}}\!\bigl(\mathcal{L}^m_{\mathrm{diff}} + \lambda_e\,\mathcal{L}^m_{\mathrm{align}}\bigr) + \lambda_{\mathrm{div}}\,\mathcal{L}_{\mathrm{div}} + \lambda_{\mathrm{reg}}\,\|\Theta\|_2^2.
    \label{eq:total_loss}
\end{equation}
Here $\mathcal{L}_{\mathrm{BPR}}$ is the pairwise ranking loss of BPR~\cite{rendle2009bpr}, computed from the GCN-fused embeddings; $\mathcal{L}_{\mathrm{CL}}$ is the multi-view contrastive loss inherited from DiffMM~\cite{10.1145/3664647.3681498}, which contrasts modality-specific user and item views against an aggregated view that does not pass through the router; $\mathcal{L}^m_{\mathrm{diff}}$ and $\mathcal{L}^m_{\mathrm{align}}$ are given in Equations~\eqref{eq:time_reweight} and~\eqref{eq:align}; $\mathcal{L}_{\mathrm{div}}$ is the router diversity regularizer (Equation~\eqref{eq:div_loss}); and $\Theta$ collects all trainable parameters. The hyperparameters $\lambda_{\mathrm{cl}}$, $\lambda_e$, $\lambda_{\mathrm{div}}$, $\lambda_t$, and $\lambda_{\mathrm{reg}}$ are tuned per dataset (see the released code for full values).

Each epoch proceeds in three phases. \textbf{(1)~Diffusion training} updates only the denoiser-side parameters $\{\theta_m\}_m$ via $\sum_m (\mathcal{L}^m_{\mathrm{diff}} + \lambda_e \mathcal{L}^m_{\mathrm{align}})$, isolating graph generation from ranking optimization. \textbf{(2)~Graph reconstruction} samples $\hat{\mathbf{x}}_0^m$ with the updated denoisers and assembles $\{\widehat{\mathbf{R}}^m\}_{m \in \mathcal{M}}$ via the top-$k$ procedure of Section~\ref{sec:reweight}. \textbf{(3)~GCN and recommendation training} fuses embeddings via the modal router and updates the remaining parameters via $\mathcal{L}_{\mathrm{BPR}} + \lambda_{\mathrm{cl}}\mathcal{L}_{\mathrm{CL}} + \lambda_{\mathrm{div}}\mathcal{L}_{\mathrm{div}} + \lambda_{\mathrm{reg}}\|\Theta\|_2^2$. Although $\mathcal{L}_{\mathrm{CL}}$ appears in Phase~3, its contrastive views are built by per-modality propagation without routing, and thus no gradient flows through the router; this prevents the contrastive objective, which can favor near-uniform weights, from pulling the router toward a collapsed global mixture.

\section{Experiments}
\label{sec:experiments}

We evaluate TimeRoute on three widely used multi-modal benchmarks: TikTok, Amazon-Baby, and Amazon-Sports.\footnote{\url{https://github.com/HKUDS/DiffMM/tree/main/Datasets}} Table~\ref{tab:dataset_stats_main} summarizes their statistics: the three datasets span different temporal ranges, timestamp coverages, and user activity distributions, which allows us to study modality time-scale mismatch under diverse settings. Full preprocessing, feature, and split details are provided in the Appendix.\footnote{\url{https://anonymous.4open.science/r/TimeRoute/TimeRoute_appendix.pdf}} We organize the experiments around four research questions:

\begin{table}[t]
\caption{Dataset and timestamp statistics. TimeCov is the fraction of interactions carrying timestamps; Range is the time span of training interactions.}
\label{tab:dataset_stats_main}
\centering
\small
\setlength{\tabcolsep}{4pt}
\begin{tabular}{@{}lrrr@{}}
\toprule
 & \textbf{TikTok} & \textbf{Amazon-Baby} & \textbf{Amazon-Sports} \\
\midrule
\# Users & 9{,}319 & 19{,}445 & 35{,}598 \\
\# Items & 6{,}710 & 7{,}050 & 18{,}357 \\
\# Interactions & 59{,}541 & 139{,}110 & 256{,}308 \\
Sparsity (\%) & 99.904 & 99.899 & 99.961 \\
TimeCov & 0.907 & 0.725 & 0.635 \\
Range & 2022.01--2023.09 & 2001.02--2014.07 & 2006.01--2023.10 \\
\bottomrule
\end{tabular}
\end{table}

\begin{itemize}[leftmargin=*]
    \item \textbf{RQ1 (Overall Effectiveness).} Does TimeRoute consistently improve top-$K$ recommendation over strong baselines across datasets and metrics?
    \item \textbf{RQ2 (Gain Attribution).} Do the gains stem merely from access to timestamp-derived inputs, or from how TimeRoute's mechanisms exploit them?
    \item \textbf{RQ3 (User-Level Routing).} Does the modal router learn systematic per-user differentiation in modality weights, and how do these weights vary with users' activity levels?
    \item \textbf{RQ4 (Component Contributions and Design Validation).} What is the individual contribution of each component, and are our architectural choices, such as restricting routing to the user side, supported empirically?
\end{itemize}

\subsection{Experimental Setup}
\label{sec:exp_setup}

\subsubsection{Evaluation metrics.}
We report three standard top-$K$ ranking metrics at $K=20$: Recall@20 (R@20), Precision@20 (P@20), and NDCG@20 (N@20); full metric definitions and dataset preprocessing are in the appendix.\footnote{\url{https://anonymous.4open.science/r/TimeRoute/TimeRoute_appendix.pdf}} Each results table additionally reports $\Delta(\%)$, the relative improvement of TimeRoute over the strongest baseline in the corresponding column.

\subsubsection{Multi-seed protocol.}
Every main experiment, including the attribution arms and all ablations, is run with 10 seeds (1, 7, 42, 123, 456, 666, 789, 888, 999, 2026); screening studies with fewer seeds are labeled as such. Attribution and ablation results (Sections~\ref{sec:rq2} and~\ref{sec:rq4}) report seed-averaged means with standard deviations.

\subsubsection{Data splits and temporal-feature protocol.}
We use a random train/validation/test split, matching prior diffusion-based multi-modal recommenders~\cite{10.1145/3664647.3681498}. A complementary chronological-split evaluation, in which each user's training interactions strictly precede held-out ones, is reported in the Appendix. \textbf{All temporal features are constructed strictly from training-set interactions. Evaluation never accesses validation or test timestamps. All model checkpoints are selected by validation Recall@20.} Thus, no validation or test timestamp influences profile construction, graph reconstruction, or model selection. Consistent with the non-sequential formulation (Section~\ref{sec:prelim}), TimeRoute does not condition on a query timestamp at prediction time; a preliminary 3-seed study found no significant paired differences with such conditioning enabled versus disabled (Amazon-Baby $p=0.38$, Amazon-Sports $p=0.60$, TikTok $p=0.43$), the main configuration therefore omits it. All experiments can be reproduced from our released code at \url{https://anonymous.4open.science/r/TimeRoute}.

\subsubsection{Compared Baselines.}
We start from the strong baseline suite used in DiffMM~\cite{10.1145/3664647.3681498} for comparability and add KDiffE~\cite{11175523}, a recent diffusion-based multimodal baseline. The comparison covers five paradigms: \textbf{(i) Conventional CF:} MF-BPR~\cite{rendle2009bpr}.
\textbf{(ii) GNN-based CF:} NGCF~\cite{wang2019neural}, LightGCN~\cite{he2020lightgcn}.
\textbf{(iii) Self-supervised CF:} SGL~\cite{wu2021self}, NCL~\cite{lin2022improving}, HCCF~\cite{xia2022hypergraph}.
\textbf{(iv) Multimodal recommenders:} VBPR~\cite{he2016vbpr}, LGCN-M~\cite{he2020lightgcn}, BM3~\cite{10.1145/3543507.3583251}, GRCN~\cite{10.1145/3394171.3413556}, MMGCN~\cite{10.1145/3343031.3351034}, LATTICE~\cite{10.1145/3474085.3475259}, CLCRec~\cite{wei2021contrastive}, MMGCL~\cite{yi2022multi}, SLMRec~\cite{tao2022self}.
\textbf{(v) Diffusion-based multimodal recommenders:} DiffMM~\cite{10.1145/3664647.3681498}, KDiffE~\cite{11175523}.
For every baseline, we follow the hyperparameter search ranges in the original paper or the public implementation. Sequential recommenders such as TiSASRec~\cite{10.1145/3336191.3371786} are excluded here: under a random split, training data contains interactions that postdate test interactions, rendering their next-item objective incompatible with this evaluation protocol. We instead compare against TiSASRec under the chronological split in the Appendix, where sequential supervision is well-defined.

\subsection{Overall Effectiveness (RQ1)}
\label{sec:rq1}

We first examine whether TimeRoute improves top-$K$ recommendation across datasets and metrics under the protocol of Section~\ref{sec:exp_setup}. Table~\ref{tab:overall_random_split} reports the comparison against all baselines.

TimeRoute outperforms every baseline on every dataset and metric. Relative gains over the strongest baseline range from $+4.92\%$ to $+9.80\%$: the largest improvements appear on P@20 and N@20 for Amazon-Baby ($+9.80\%$ and $+9.73\%$) and on N@20 for Amazon-Sports and TikTok ($+7.83\%$ and $+7.17\%$), indicating improvements in precision and ranking quality, not only recall.

DiffMM and KDiffE are the strongest competitors on all three datasets; TimeRoute improves over both while sharing their backbone family, suggesting that the margin reflects the temporal components rather than a fundamentally different modeling paradigm. TimeRoute also differs from all baselines in consuming timestamp-derived inputs, and the margin could in principle reflect information access rather than mechanism design; RQ2 addresses this with controlled comparisons.

\begin{table*}[thb]
\caption[Overall performance on the random split]{Overall performance on the \textit{Random split} (R@20, P@20, N@20). $\Delta(\%)$ is the relative improvement of TimeRoute (mean over 10 seeds) over the best baseline per column. Best in \textbf{bold}; second-best \underline{underlined}.}
\label{tab:overall_random_split}
\centering
\small
\setlength{\tabcolsep}{4pt}
\begin{tabular}{@{}lrrrrrrrrr@{}}
\toprule
 & \multicolumn{3}{c}{\textbf{TikTok}} & \multicolumn{3}{c}{\textbf{Amazon-Baby}} & \multicolumn{3}{c}{\textbf{Amazon-Sports}} \\
\cmidrule(lr){2-4} \cmidrule(lr){5-7} \cmidrule(lr){8-10}
 & \textbf{R@20} & \textbf{P@20} & \textbf{N@20} & \textbf{R@20} & \textbf{P@20} & \textbf{N@20} & \textbf{R@20} & \textbf{P@20} & \textbf{N@20} \\
\midrule
\multicolumn{10}{@{}l}{\textit{General collaborative filtering}} \\
\quad MF-BPR~\cite{rendle2009bpr}        & 0.0346 & 0.0017 & 0.0130 & 0.0440 & 0.0024 & 0.0200 & 0.0430 & 0.0023 & 0.0202 \\
\quad NGCF~\cite{wang2019neural}         & 0.0604 & 0.0030 & 0.0238 & 0.0591 & 0.0032 & 0.0261 & 0.0695 & 0.0037 & 0.0318 \\
\quad LightGCN~\cite{he2020lightgcn}     & 0.0653 & 0.0033 & 0.0282 & 0.0698 & 0.0037 & 0.0319 & 0.0782 & 0.0042 & 0.0369 \\
\quad SGL~\cite{wu2021self}              & 0.0603 & 0.0030 & 0.0238 & 0.0678 & 0.0036 & 0.0296 & 0.0779 & 0.0041 & 0.0361 \\
\quad NCL~\cite{lin2022improving}        & 0.0658 & 0.0034 & 0.0269 & 0.0703 & 0.0038 & 0.0311 & 0.0765 & 0.0040 & 0.0349 \\
\quad HCCF~\cite{xia2022hypergraph}      & 0.0662 & 0.0029 & 0.0267 & 0.0705 & 0.0037 & 0.0308 & 0.0779 & 0.0041 & 0.0361 \\
\cmidrule(lr){1-10}
\multicolumn{10}{@{}l}{\textit{Multi-modal recommendation}} \\
\quad VBPR~\cite{he2016vbpr}             & 0.0380 & 0.0018 & 0.0134 & 0.0486 & 0.0026 & 0.0213 & 0.0582 & 0.0031 & 0.0265 \\
\quad LGCN-M~\cite{he2020lightgcn}       & 0.0679 & 0.0034 & 0.0273 & 0.0726 & 0.0038 & 0.0329 & 0.0705 & 0.0035 & 0.0324 \\
\quad MMGCN~\cite{10.1145/3343031.3351034}        & 0.0730 & 0.0036 & 0.0307 & 0.0640 & 0.0032 & 0.0284 & 0.0638 & 0.0034 & 0.0279 \\
\quad GRCN~\cite{10.1145/3394171.3413556}         & 0.0804 & 0.0036 & 0.0350 & 0.0754 & 0.0040 & 0.0336 & 0.0833 & 0.0044 & 0.0377 \\
\quad LATTICE~\cite{10.1145/3474085.3475259}      & 0.0843 & 0.0042 & 0.0367 & 0.0829 & 0.0044 & 0.0368 & 0.0915 & 0.0048 & 0.0424 \\
\quad CLCRec~\cite{wei2021contrastive}   & 0.0621 & 0.0032 & 0.0264 & 0.0610 & 0.0032 & 0.0284 & 0.0651 & 0.0035 & 0.0301 \\
\quad MMGCL~\cite{yi2022multi}           & 0.0799 & 0.0037 & 0.0326 & 0.0758 & 0.0041 & 0.0331 & 0.0875 & 0.0046 & 0.0409 \\
\quad SLMRec~\cite{tao2022self}          & 0.0845 & 0.0042 & 0.0353 & 0.0765 & 0.0043 & 0.0325 & 0.0829 & 0.0043 & 0.0376 \\
\quad BM3~\cite{10.1145/3543507.3583251}       & 0.0957 & 0.0048 & 0.0404 & 0.0839 & 0.0044 & 0.0361 & 0.0975 & 0.0051 & 0.0442 \\
\quad DiffMM~\cite{10.1145/3664647.3681498}       & \underline{0.1129} & 0.0056 & 0.0456 & \underline{0.0975} & \underline{0.0051} & \underline{0.0411} & 0.1017 & 0.0054 & 0.0458 \\
\quad KDiffE~\cite{11175523}               & 0.1120 & \underline{0.0056} & \underline{0.0460} & 0.0950 & 0.0050 & 0.0400 & \underline{0.1020} & \underline{0.0056} & \underline{0.0460} \\
\midrule
\textbf{TimeRoute}      & \textbf{0.1208} & \textbf{0.0059} & \textbf{0.0493} & \textbf{0.1023} & \textbf{0.0056} & \textbf{0.0451} & \textbf{0.1094} & \textbf{0.0059} & \textbf{0.0496} \\
\textit{$\Delta$\,(\%)} & \textit{+7.00}  & \textit{+5.36}  & \textit{+7.17}  & \textit{+4.92}  & \textit{+9.80}  & \textit{+9.73}  & \textit{+7.25}  & \textit{+5.36}  & \textit{+7.83}  \\
\bottomrule
\end{tabular}
\end{table*}

\subsection{Gain Attribution (RQ2)}
\label{sec:rq2}

The gains in RQ1 might reflect access to timestamp-derived inputs rather than the proposed mechanisms. We probe this from opposite directions: the first comparison holds the backbone fixed and varies \emph{how} timestamp information enters the model; the second holds the architecture fixed and corrupts \emph{what} the router receives.

\subsubsection{Is access to timestamps alone sufficient?}
\label{sec:rq2_expb}
We compare three configurations that share the backbone, hyperparameters, and seeds. \textbf{Plain} removes all temporal components from TimeRoute: no router, no FiLM conditioning, no dual-stream heads, and no time-reweighted loss, leaving the model with no access to timestamps (the FiLM layers remain in place but receive a zero vector, reducing to a learned constant modulation that carries no temporal information). \textbf{Naive} gives the same backbone the identical per-user temporal profile, injected without any of the proposed temporal mechanisms: the profile, in the same 12-dimensional raw form consumed by the diffusion conditioner (Section~\ref{sec:prelim}), passes through a small two-layer MLP (zero-initialized at the output) whose result is added to the user context that conditions the denoiser. \textbf{Full} is TimeRoute. If access to timestamp information were sufficient to explain the gains, Naive should recover most of Full's advantage over Plain.

\begin{table}[thb]
\caption[Three-configuration attribution study]{Three-configuration attribution study (R@20, mean$\pm$std over 10 seeds). Plain: backbone with all temporal components removed (no timestamp access). Naive: the same per-user temporal profile injected without the proposed mechanisms. Full: TimeRoute. Best results per column in \textbf{bold}.}
\label{tab:expb_threearm}
\centering
\footnotesize
\setlength{\tabcolsep}{4pt}
\begin{tabular}{@{}lccc@{}}
\toprule
\textbf{Configuration} & \textbf{Amazon-Baby} & \textbf{Amazon-Sports} & \textbf{TikTok} \\
\midrule
Plain (no temporal input) & $0.0987{\scriptstyle\pm0.0009}$ & $0.1060{\scriptstyle\pm0.0012}$ & $0.1122{\scriptstyle\pm0.0023}$ \\
Naive injection           & $0.0963{\scriptstyle\pm0.0009}$ & $0.1059{\scriptstyle\pm0.0006}$ & $0.1108{\scriptstyle\pm0.0021}$ \\
TimeRoute (full)          & $\mathbf{0.1023{\scriptstyle\pm0.0005}}$ & $\mathbf{0.1094{\scriptstyle\pm0.0007}}$ & $\mathbf{0.1208{\scriptstyle\pm0.0009}}$ \\
\bottomrule
\end{tabular}
\end{table}

Table~\ref{tab:expb_threearm} shows that naive access alone does not help. Naive never improves over Plain on any dataset: it is numerically worse on Amazon-Baby ($\Delta=-0.0024$) and TikTok ($\Delta=-0.0014$), and nearly identical to Plain on Amazon-Sports ($\Delta=-0.0001$). Injecting the very same temporal profile through an unconstrained pathway thus yields no benefit and can even hurt, suggesting that directly perturbing the user context may distort representations that the ranking objective depends on. In contrast, Full outperforms both Plain and Naive on all three datasets by margins that substantially exceed the reported seed-level standard deviations, with gains over Plain of $+0.0036$ (Amazon-Baby), $+0.0034$ (Amazon-Sports), and $+0.0086$ (TikTok). The same information that provides no gain when injected naively becomes consistently beneficial when consumed through the proposed pathway: the gains in Table~\ref{tab:overall_random_split} therefore depend on how timestamp-derived information is incorporated, not merely on its availability.

\subsubsection{Is temporal input necessary?}
\label{sec:rq2_noise}
The second comparison probes the mechanism from the opposite direction on Amazon-Baby. We keep the full architecture and replace the router's temporal input with fixed random noise of identical dimensionality (a 16-dimensional Gaussian vector per user, drawn with a seed independent of the training seed), leaving the parameter count unchanged. This preserves the router's capacity for per-user adaptive weighting while removing the temporal content of its input. We additionally compare against ablating the router entirely.

The result shows that the router's benefit depends on temporal input. Replacing the temporal profile with noise drops R@20 from $0.1023{\scriptstyle\pm0.0005}$ to $0.0984{\scriptstyle\pm0.0004}$ (mean$\pm$std over 10 seeds), a $3.8\%$ degradation that substantially exceeds the reported seed-level standard deviations: the router cannot recover its benefit from an input that carries no temporal information, even though its architecture, capacity, and parameter count are unchanged. Moreover, the noise-fed router performs no better than removing the router entirely ($0.0984$ vs.\ $0.0985{\scriptstyle\pm0.0006}$): per-user adaptive weighting alone provides no measurable benefit in this probe, indicating that the router's value lies in the temporal content of its input rather than its added capacity. We treat this single-dataset probe as a mechanism check rather than a claim about every dataset.

Together, the two comparisons show that temporal information without the proposed mechanisms provides no gain, and that the router's benefit depends on temporal input: TimeRoute's gains arise from the \emph{conjunction} of the routing pathway and temporal input that encodes each user's behavior.

\subsection{User-Level Routing (RQ3)}
\label{sec:rq3}

Does the router learn \emph{differentiated} per-user weights, or does it converge to a single global weighting that merely improves on DiffMM's shared default? If all users receive nearly identical weights, the router reduces to a learned global mixture and provides no user-specific adaptation. To investigate, we partition users on Amazon-Baby and TikTok into four interaction-count quartiles, where Q1 contains the least active users and Q4 the most active. Because interaction counts are highly discrete, quartile boundaries fall on values with many ties and the resulting groups need not be equally sized; on TikTok in particular, Q4 contains a small set of exceptionally active users (the full per-quartile statistics, including group sizes, interaction-count ranges, and within-quartile weight dispersion, are listed in the Appendix). Figure~\ref{fig:user_routing_quartiles} plots, for each quartile, the mean modality weights $w_{\text{img}}$ and $w_{\text{txt}}$ ($w_{\text{aud}}$ additionally on TikTok) together with the mean per-user routing entropy $H = \mathbb{E}_u[\mathrm{H}(\mathbf{w}_u)]$ and the number of users $n$.

\begin{figure}[thb]
\centering
\begin{subfigure}[b]{0.48\linewidth}
\includegraphics[width=\linewidth]{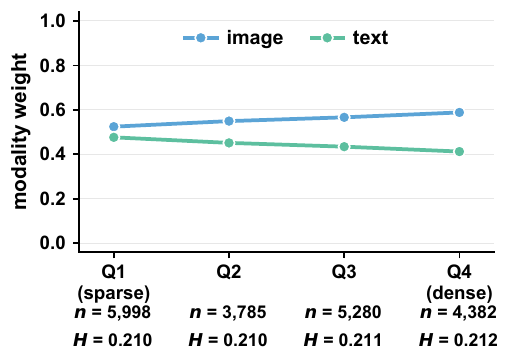}
\caption{Amazon-Baby ($N = 19{,}445$).}
\label{fig:user_routing_quartiles_baby}
\end{subfigure}
\hfill
\begin{subfigure}[b]{0.48\linewidth}
\includegraphics[width=\linewidth]{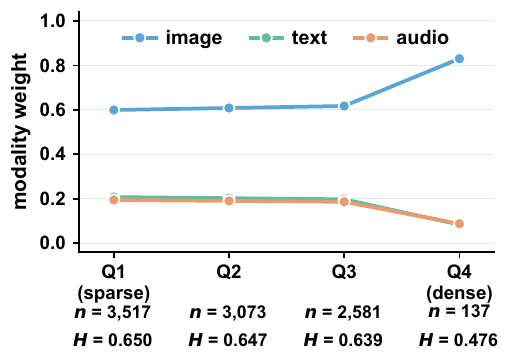}
\caption{TikTok ($N = 9{,}308$).}
\label{fig:user_routing_quartiles_tiktok}
\end{subfigure}
\caption[Per-user routing analysis by activity quartile]{Per-user routing analysis by user activity quartile on (a) Amazon-Baby and (b) TikTok. For each quartile, we plot the mean modality weights $w_{\text{img}}$, $w_{\text{txt}}$ (and $w_{\text{aud}}$ for TikTok) together with the mean per-user routing entropy. Q1 contains users with the fewest interactions; Q4 contains the most active users. On TikTok, $N=9{,}308$ denotes the users present in the released timestamped interaction files, slightly below the benchmark's $9{,}319$ users in Table~\ref{tab:dataset_stats_main}.}
\label{fig:user_routing_quartiles}
\Description{Two line charts showing per-user modality routing weights across user activity quartiles Q1 to Q4 on Amazon-Baby and TikTok.}
\end{figure}

Figure~\ref{fig:user_routing_quartiles} shows two patterns. First, the router differentiates at the level of individual users rather than activity bands. On Amazon-Baby, the mean image weight rises from $0.524$ at Q1 to $0.588$ at Q4 ($+12.2\%$ relative) while the mean per-user routing entropy remains nearly constant ($0.210$ to $0.212$): higher activity is associated with a shift in \emph{where} weight is allocated, not in \emph{how confidently}. The weights also vary widely within each quartile (image-weight dispersion $0.436$--$0.445$; see the Appendix), and thus under a two-modality simplex many users sit near one extreme or the other; combined with the low mean entropy relative to the two-modality bound ($\ln 2 \approx 0.693$), this indicates confident, user-specific routing rather than a global solution with minor perturbations, consistent with the behavior the diversity regularizer (Section~\ref{sec:diversity}) is designed to produce.

Second, the differentiation is directional on TikTok. The mean image weight climbs from $0.599$ at Q1 to $0.617$ at Q3, then jumps to $0.830$ for the most active quartile, while text and audio fall to $0.083$ and $0.087$, mean entropy drops from $0.650$ to $0.476$, and within-quartile dispersion shrinks from $0.363$ to $0.194$. Q4 is a small set of extremely active users ($n=137$, reflecting the platform's heavy-tailed activity): higher activity is associated with progressively stronger commitment to image, the dominant learned modality, while the weight floor (Section~\ref{sec:floor}) bounds the rest. This specialization occurs within a router that the ablation in Section~\ref{sec:rq4} shows to be beneficial overall (removing the router costs $3.56\%$ of R@20 on TikTok); we note it as a boundary of adaptive fusion and return to it in the Limitations. The quartile trends alone do not show that \emph{temporal} content specifically, rather than activity level, structures the weights; that evidence comes from the noise-input probe of Section~\ref{sec:rq2_noise}.

\subsection{Component Contributions and Design Validation (RQ4)}
\label{sec:rq4}

We now ask which architectural components contribute to these gains. Four ablations each remove one component while keeping the rest intact: w/o Modal Router, w/o Dual-Stream, w/o FiLM, and w/o Time-Weight. Every ablation follows the full 10-seed protocol of Section~\ref{sec:exp_setup}, and Table~\ref{tab:ablation_random} reports R@20 for each variant together with its relative change with respect to the full model.

\begin{table}[thb]
\caption[Component ablation on the random split]{Ablation study on the Random split. Each row removes one component from the full model. We report R@20 as mean $\pm$ std over 10 seeds and the relative change with respect to the full model.}
\label{tab:ablation_random}
\centering
\scriptsize
\setlength{\tabcolsep}{1pt}
\begin{tabular}{@{}lccc@{}}
\toprule
\textbf{Setting} & \textbf{Baby} & \textbf{Sports} & \textbf{TikTok} \\
\midrule
\textbf{Full Model}
 & $\mathbf{0.1023 \pm 0.0005}$
 & $\mathbf{0.1094 \pm 0.0007}$
 & $\mathbf{0.1208 \pm 0.0009}$ \\
w/o Modal Router
 & $0.0985 \pm 0.0006$ \textit{($-$3.71\%)}
 & $0.1062 \pm 0.0008$ \textit{($-$2.93\%)}
 & $0.1165 \pm 0.0010$ \textit{($-$3.56\%)} \\
w/o Dual-Stream
 & $0.0994 \pm 0.0007$ \textit{($-$2.83\%)}
 & $0.1058 \pm 0.0006$ \textit{($-$3.29\%)}
 & $0.1159 \pm 0.0008$ \textit{($-$4.06\%)} \\
w/o FiLM
 & $0.0993 \pm 0.0005$ \textit{($-$2.93\%)}
 & $0.1070 \pm 0.0007$ \textit{($-$2.19\%)}
 & $0.1173 \pm 0.0009$ \textit{($-$2.90\%)} \\
w/o Time-Weight
 & $0.0986 \pm 0.0006$ \textit{($-$3.62\%)}
 & $0.1064 \pm 0.0008$ \textit{($-$2.74\%)}
 & $0.1165 \pm 0.0010$ \textit{($-$3.56\%)} \\
\bottomrule
\end{tabular}
\end{table}

\noindent $\bullet$ \textbf{Every temporal component is individually necessary.} Removing any single component consistently degrades R@20 on all three datasets, with relative drops between $2.19\%$ and $4.06\%$. The degradation spans both of TimeRoute's pathways: the fusion-level modal router (Problem~1) and each of the three diffusion-side temporal components (Problem~2). Temporal awareness thus contributes at both levels of the framework, rather than being concentrated in a single module.

\noindent $\bullet$ \textbf{The modal router contributes at the fusion level.} Removing the router reduces R@20 by $3.71\%$ on Amazon-Baby, $2.93\%$ on Amazon-Sports, and $3.56\%$ on TikTok. Notably, the router remains beneficial on Amazon-Sports despite that dataset's partial timestamp coverage (approximately 63\% of training interactions), suggesting that the routing pathway can remain effective under incomplete temporal observability: profiles built from the timestamped subset of a user's history still provide a usable routing signal.

\noindent $\bullet$ \textbf{The diffusion-side components contribute at the reconstruction level.} Dual-Stream, FiLM, and Time-Weight each make a consistent contribution: removing the dual-stream heads produces the largest single drop on TikTok ($-4.06\%$) and Amazon-Sports ($-3.29\%$), removing the time-reweighted loss reduces R@20 by $3.62\%$ on Amazon-Baby, and removing FiLM conditioning reduces R@20 by between $2.19\%$ and $2.93\%$ across datasets. This shows that time-conditioned graph reconstruction (Problem~2) makes a distinct, non-redundant contribution in addition to fusion-level routing: even with the router intact, removing any one of these components reduces performance. That the dual-stream separation of long- and short-term dynamics shows its largest effect on TikTok is consistent with the platform's fast-moving interaction patterns, which may make the separation of long- and short-term signals particularly useful.

\noindent $\bullet$ \textbf{User-side versus item-side routing.} As an additional check on the fusion design of Section~\ref{sec:fusion}, we evaluated an item-side router variant that builds per-item temporal profiles and applies analogous routing to item-side fusion; it consistently reduced R@20 on all three datasets, and the full setup and numbers are reported in the Appendix.

\section{Conclusion}

We proposed \textbf{TimeRoute}, a diffusion-based multi-modal recommender for modality time-scale mismatch, combining a temporal-aware modal router that personalizes modality fusion per user with time-conditioned diffusion reconstruction that suppresses outdated graph-level signals. Across TikTok, Amazon-Baby, and Amazon-Sports, experiments averaged over 10 seeds show consistent improvements over strong baselines. Controlled attribution studies show that the gains require both the proposed mechanisms and temporal input; quartile analysis finds the learned routing to be systematically user-specific; and component ablations show every tested temporal component contributes consistently.

\section*{Limitations}
First, routing is static per user: each user receives one fixed fusion distribution aggregated from their training history; per-moment dynamic routing is a natural next step. Second, our ablations are one-component-at-a-time and the noise-input probe covers Amazon-Baby only; interaction effects between the two pathways, which the Plain-arm comparison suggests are partly synergistic, remain unquantified. Third, the main evaluation follows the random-split protocol of prior work, and the chronological-split evaluation in the Appendix covers fewer baselines (DiffMM and TiSASRec). Fourth, routing quality may degrade under partial timestamp coverage. Finally, on TikTok, highly active users receive near-single-modality weights, increasing reliance on the dominant modality. Future work could add modality-quality estimation and extend TimeRoute to sequential recommendation.

\section*{Generative AI Disclosure}
We used a large language model (LLM) to assist with English proofreading and minor grammar edits in parts of the manuscript.

\section*{Ethics and Privacy Statement}
This work uses only publicly available benchmark datasets about public entities, and involves no human subjects or personal data, we therefore foresee minimal direct risk. We release our code to support transparency and responsible use.



\bibliographystyle{ACM-Reference-Format}
\bibliography{sample-base}


\end{document}